\documentclass[aps,prd,12pt,amssymb]{revtex4}
                                                                          
\begin{document}                                                                         
                                                                          
\baselineskip=0.60cm
                                                                          
\newcommand{\ini}{\begin{equation}}
\newcommand{\fin}{\end{equation}}
\newcommand{\inir}{\begin{eqnarray}}
\newcommand{\finr}{\end{eqnarray}}
                                                                                
\def\ol{\overline}
\def\pa{\partial}
\def\ra{\rightarrow}
\def\ts{\times}
\def\df{\dotfill}
\def\bs{\backslash}
\def\dg{\dagger}
\def\la{\lambda}
\def\ep{\epsilon}
                                                                                
$~$
      
\vspace{1 cm}
                                                                               
\title{Leptogenesis within a generalized quark-lepton symmetry}

\author{F. Buccella}

\affiliation{Dipartimento di Scienze Fisiche,
Universit\`a di Napoli, Via Cintia, Napoli, Italy}
\affiliation{INFN, Sezione di Napoli, Italy}

\author{D. Falcone}

\affiliation{Dipartimento di Scienze Fisiche,
Universit\`a di Napoli, Via Cintia, Napoli, Italy}

\author{L. Oliver}

\affiliation{Laboratoire de Physique Theorique, Universit\'e de Paris XI,
Batiment 210, Orsay Cedex, France}

\begin{abstract}
\vspace{1cm}
\noindent
Quark-lepton symmetry has been shown to be inconsistent with baryogenesis 
via leptogenesis in natural schemes of the see-saw mechanism. Within the
phenomenological approach of textures, we relax this strict symmetry and
propose weaker conditions, namely models of the neutrino Dirac mass matrix
$M_D$ which have the same hierarchy as the matrix elements of $M_u$. We call
this guide-line generalized $hierarchical$ quark-lepton symmetry. We consider
in detail particular cases in which the moduli of the matrix elements of
$M_D$ are equal to those of $M_u$. Within the phenomenological approach of
textures, we try for the heavy Majorana mass matrix diagonal and off-diagonal
forms. We find that an ansatz for $M_D$ preserving the hierarchy, together
with an off-diagonal model for the heavy Majorana neutrino mass, is consistent 
with neutrino masses, neutrino mixing and baryogenesis via leptogenesis for an 
intermediate mass scale $m_R \sim 10^{12}$ GeV. The preservation of the
hierarchical structure could come from a possible symmetry scheme.
\end{abstract}
                                                                                
\maketitle
                                                                                
\newpage

\section{Introduction}

The seesaw mechanism \cite{ss} can explain the smallness of neutrino 
masses and is consistent with large lepton mixing \cite{lm}. In a nutshell, it is based 
on the existence of very heavy right-handed Majorana neutrinos.
A cosmological consequence is the generation of a baryon asymmetry in the universe by means of the 
out-of-equilibrium decays of the heavy right-handed neutrinos to leptons
and $SU(2)_L$ Higgs bosons, which create a lepton asymmetry, partially converted to a baryon
asymmetry by electroweak sphalerons, a mechanism known as baryogenesis via
leptogenesis \cite{fy}.

Recently, much progress has been done in the study 
of leptogenesis, especially in the exploration of flavor effects
\cite{abada}. In the present paper we turn again to the link between
leptogenesis and fermion  mass matrices. We are interested in the 
compatibility of quark-lepton symmetry with leptogenesis. In previous papers
this compatibility has been strongly questioned \cite{no}. In Ref. \cite{afs} it is achieved only in the case of some degeneracy of the
right-handed neutrino masses. This was also argued in Ref. \cite{bft}.
Often an inverse seesaw formula was used. In order to further explore this 
subject, we adopt here a typical form of the quark mass matrices 
\cite{akm}, together with minimal models for the Majorana mass
matrix \cite{df1}, and the direct seesaw formula. In a second step we relax quark-lepton symmetry, and adopt a weaker hypothesis, namely
keeping the hierarchy of the Dirac mass matrix elements, with the moduli of the elements of $M_D$ being equal to the moduli of the elements of $M_u$.

\section{Mass matrices}

According to the seesaw mechanism, the effective mass matrix of neutrinos
is given by the formula
\ini
M_{\nu} \simeq M_D M_R^{-1} M_D,
\fin
where $M_D$ is the Dirac mass matrix and $M_R$ the Majorana mass matrix.
For $M_R \gg M_D$, we have $M_{\nu} \ll M_D$.

Our starting point for fermion mass matrices is the following symmetric forms of 
quark mass matrices \cite{akm,beg}, for which we simply give an order of
magnitude of the matrix elements,
\ini
M_u \simeq
\left( \begin{array}{ccc}
0 & i \ep{_u^3} & \ep{_u^4} \\
i \ep{_u^3} & \ep{_u^2} & \ep{_u^2} \\
\ep{_u^4} &  \ep{_u^2} & 1
\end{array} \right)m_t,
\fin
\ini
M_d \simeq
\left( \begin{array}{ccc}
0 & \ep{_d^3} & \ep{_d^4} \\
\ep{_d^3} & \ep{_d^2} & \ep{_d^2} \\
\ep{_d^4} & \ep{_d^2} & 1
\end{array} \right)m_b,
\fin
with $\ep_u^2 = m_c/m_t$, $\ep_d^2 = m_s/m_b$, i.e. $\ep_u \simeq 0.05$ and $\ep_d \simeq 0.15$, which agree with the 
mass spectrum of the quarks and the CKM mixing matrix.
Then, simple quark-lepton symmetry leads to the relations
\ini
M_D = M_u,~M_e = M_d
\fin
for example in $SO(10)$ with Higgses transforming as {\bf 10} 
representations. Indeed, $SO(10)$ is a favored scenario for neutrino
mass and leptogenesis, since one has right-handed neutrinos with
heavy Majorana masses and $B-L$ spontaneous symmetry breaking.
As it is well-known, the relation 
$M_e = M_d$ can be naturally modified in SO(10) with a {\bf 126}
representation in order to have $-3$ Clebsch-Gordan coefficient, coming 
from color, in the (2,2) entry for $M_e$ relatively to $M_d$, that yields
the better relation $m_s = m_{\mu}/3$ at the unification scale \cite{gj}.

For the right-handed neutrino we take minimal mass matrices, namely :

(i)  the 
diagonal,
\ini
M_R =
\left( \begin{array}{ccc}
a & 0 & 0 \\
0 & b & 0 \\
0 & 0 & 1
\end{array} \right)m_R,
\fin
and (ii) the off-diagonal
\ini
M_R =
\left( \begin{array}{ccc}
0 & 0 & 1 \\
0 & c & 0 \\
1 & 0 & 0
\end{array} \right)m_R.
\fin

Moreover, in the diagonal model we will choose $b \simeq \ep^4$, $a \simeq \ep^5$,
and in the off-diagonal model $c \simeq \ep^2$, as explained below. We also examine other cases of rank-3 matrices in the Appendix.

In the two cases (5) and (6), application of the seesaw formula gives the 
phenomenological viable form \cite{vis,moro}
\ini
M_{\nu} \sim
\left( \begin{array}{ccc}
\ep^2 & \ep & \ep \\
\ep & 1 & 1 \\
\ep & 1 & 1
\end{array} \right)m_{\nu},
\fin
where, from now on, we denote 
\ini
\ep = \ep_u \simeq 0.05
\fin

This neutrino mass matrix corresponds to maximal mixing and a normal hierarchy with the overall scale
$m_{\nu} \simeq 0.05$ eV, fixed by the atmospheric and accelerator 
neutrino 
oscillations, but with two different mass scales $m_R$.\par
The matrix (7) gives only a qualitative account for the experimental situation
of neutrino masses and mixing, since it yields the square mass differences,
$m{_3^2} - m{_2^2} \sim (0.1~eV)^2$, $m{_2^2} - m{_1^2} \sim 0$ and,
after diagonalization of $M_{\nu}$ and $M_e$, a large lepton mixing.\par
It must be emphasized that our purpose in this paper is only an 
order-of-magnitude analysis, namely to examine the consistency between a 
neutrino spectrum and a lepton mixing matrix with approximately maximal mixing,
and the amount of needed leptogenesis to explain the baryon asymmetry of the
universe.

\section{Leptogenesis}

Since we are interested in an order-of-magnitude calculation, 
we consider leptogenesis formulas in the single-flavor approximation 
\cite{abada}. The calculation must be done in the basis where the right-handed
mass matrix is diagonal (with eigenvalues $M_1, M_2, M_3$).
The baryon asymmetry, baryon to entropy fraction, is given by
\ini
Y_B \simeq \frac{1}{2} Y_L
\fin
and the lepton asymmetry by
\ini
Y_L \simeq 0.3 ~\frac{\epsilon_1}{g_*}
\left( \frac{0.55 \cdot 10^{-3} eV}{\tilde{m}_1} \right)^{1.16}
\fin
in the strong washout regime, and
\ini
Y_L \simeq 0.3 ~\frac{\epsilon_1}{g_*}
\left( \frac{\tilde{m}_1}{3.3 \cdot 10^{-3}eV} \right)
\fin
in the opposite weak washout regime. The parameter  $g_*$ is the number of light degrees of freedom, of the order $g_* \simeq 100$ in the standard case. 
Strong washout is realized for $\tilde{m}_1 \gg 3 \cdot 10^{-3}$,
where $\tilde{m}_1 = (M_D^{\dg} M_D)_{11} /M_1$. \par Notice that $Y_B$ is smaller than the baryon to photon ratio $\eta$ by roughly a factor 7. The experimental value of the baryon asymmetry is (see \cite{wmap}),
\ini
(Y_B)_{exp} \simeq 9 \cdot 10^{-11}
\fin 

The CP-violating asymmetry $\epsilon_1$,
related to the decay of the lightest right-handed neutrino is given here below case
by case. We now consider the two different textures for the right-handed neutrino mass matrices proposed above.

\section{The diagonal model}

In the diagonal model (i), application of the seesaw formula gives the 
effective neutrino mass matrix
\ini
M_{\nu} \simeq
\left( \begin{array}{ccc}
-\ep^6/b+\ep^8 & i \ep^5/b+\ep^6 & i \ep^5/b+\ep^4 \\
{*} & -\ep^6/a+\ep^4/b+\ep^4 & i \ep^7/a+\ep^4/b+\ep^2 \\
{*} & {*} & \ep^8/a+\ep^4/b+1
\end{array} \right) \frac{m_t^2}{m_R}.
\fin
A structure similar to (7) is achieved for
$b \sim \ep^4$ and $a \sim \ep^5$.\par 
Since we hopefully expect 
\ini
\frac{m_t^2}{m_R}\simeq0.05~eV 
\fin
to describe the neutrino spectrum and lepton mixing, from (7)
we get 
\ini
m_R \sim 10^{15}~GeV,
\fin
near a unification scale.
In this case
\ini
\epsilon_1 \simeq \frac{3}{16 \pi v^2} \left(
\frac{Im(M_D^{\dg}M_D)^2_{12}}{(M_D^{\dg}M_D)_{11}}
\frac{M_1}{M_2}+
\frac{Im(M_D^{\dg}M_D)^2_{13}}{(M_D^{\dg}M_D)_{11}}
\frac{M_1}{M_3} \right),
\fin
where $v$ is the v.e.v. of the Higgs field,
and we get 

\ini
\epsilon_1 \simeq 2 \cdot 10^{-9}
\fin
We have also 
\ini
\tilde{m}_1 \simeq \ep m_{\nu},
\fin 
that lies in the weak washout regime, so that
\ini
Y_B \simeq 2 \cdot 10^{-12}
\fin

Note that $a \sim \ep^4$ is also possible. In such a case there is degeneracy 
in the lightest right-handed neutrinos and the lepton asymmetry is 
enhanced \cite{ab}. Note also that although we do not consider flavor 
effects, they can in principle preserve the asymmetry related to the decay of
the second neutrino \cite{vives}; here  $\epsilon_2 \sim 10^{-13}$.
 
\section{The off-diagonal model}
 
In the off-diagonal model (ii), the seesaw formula gives
\ini
M_{\nu} \simeq
\left( \begin{array}{ccc}
-\ep^6/c & i \ep^7+i \ep^5/c & \ep^8+i \ep^5/c \\
{*} & i \ep^5+\ep^4/c+i \ep^5 & \ep^6+\ep^4/c+i \ep^3 \\
{*} & {*} & \ep^4+\ep^4/c+\ep^4
\end{array} \right) \frac{m_t^2}{m_R}.
\fin
A structure similar to (7) is now achieved for
$\ep^4 \lesssim c \lesssim \ep^2$. For $c \simeq \ep^4$ one has
$m_{\nu} \simeq {m_t^2}/{m_R}$, while in the more interesting case $c 
\simeq \ep^2$ one has
\ini
m_{\nu} \simeq \ep^2 \frac{m_t^2}{m_R} \simeq 0.05~eV, 
\fin 
and hence, from $\ep\simeq 0.05~eV$, the intermediate scale 
\ini
m_R \sim 10^{12}~GeV
\fin
In this last case
\ini
\epsilon_1 \simeq \frac{3}{16 \pi v^2} \left(
\frac{Im(M_D^{\dg}M_D)^2_{12}}{(M_D^{\dg}M_D)_{22}}
\frac{M_2}{M_1}+
\frac{Im(M_D^{\dg}M_D)^2_{23}}{(M_D^{\dg}M_D)_{22}}
\frac{M_2}{M_3} \right).
\fin
After a right-handed rotation in the 1-3 sector, we get  for $c \simeq \ep^2$,
\ini
\epsilon_1 \simeq 5 \cdot 10^{-11}
\fin
With smaller values of the parameter $c$ we obtain a larger scale 
$m_R$ and a smaller amount of the CP asymmetry.\par
Notice one point. The simple ansatz (6) implies two degenerate very heavy Majorana neutrinos $M_1 = M_3 = m_R$ and a lighter one $M_2 \simeq \ep^2~m_R$. This is a quite different situation as the one considered in \cite{afs,bft}, that needed a quasi-degeneracy of the lightest heavy neutrinos decaying out of equilibrium. Of course, the proposal (6) can be easily modified to have three heavy Majorana neutrinos of quite different masses.   

\section{Relaxing quark-lepton symmetry preserving hierarchy of matrix elements}

From the results of the preceding Sections, we realize that keeping strict quark-lepton symmetry (4) with $M_u$ given by (2), both the diagonal and the off-diagonal models for the heavy Majorana right-handed neutrinos provide a too small baryon 
asymmetry.

Keeping the two models for the heavy Majorana right-handed neutrinos, we will now try to modify the quark-lepton symmetry relation (4), while preserving for the $M_D$ matrix elements the same order of magnitude in powers of $\ep$, i.e. we relax the quark-lepton symmetry relation while keeping the same hierarchy. Instead of $M_D = M_u$ with $M_u$ given by (2), we propose then a Dirac neutrino mass matrix of the form 

\ini
M_D \simeq
\left( \begin{array}{ccc}
0 & O(\ep^3) & O(\ep^{4}) \\
O(\ep^3) & O(\ep^2) & O(\ep^2) \\
O(\ep^{4}) & O(\ep^2) & O(1)
\end{array} \right)m_t,
\fin
The interest of such an ansatz is that possible symmetries could link this matrix to $M_u$.

To have a guide about which scheme of this kind may give the right baryon asymmetry, we perform the following exercise. We modify the Dirac mass matrix $M_D$ given by $M_D = M_u$ by putting $i$ factors in several matrix elements. We make the trial

\ini
M_D \simeq
\left( \begin{array}{ccc}
0 & \ep^3 & \ep^{4} \\
\ep^3 & \ep^2 & \ep^2 \\
\ep^{4} & \ep^2 & 1
\end{array} \right)m_t
\fin
i.e. we drop the $i$ factors in (2). Of course, this matrix is real and cannot lead to a CP asymmetry. However, this form can give us a hint of the possible interesting cases. With this last form of $M_D$ we begin by computing the $real$ quantities 

\ini
\eta_1 \simeq \frac{3}{16 \pi v^2} \left(
\frac{(M_D^{\dg}M_D)^2_{12}}{(M_D^{\dg}M_D)_{11}}
\frac{M_1}{M_2}+
\frac{(M_D^{\dg}M_D)^2_{13}}{(M_D^{\dg}M_D)_{11}}
\frac{M_1}{M_3} \right),
\fin
 for the diagonal case (i), and

\ini
\eta_1 \simeq \frac{3}{16 \pi v^2} \left(
\frac{(M_D^{\dg}M_D)^2_{12}}{(M_D^{\dg}M_D)_{22}}
\frac{M_2}{M_1}+
\frac{(M_D^{\dg}M_D)^2_{23}}{(M_D^{\dg}M_D)_{22}}
\frac{M_2}{M_3} \right).
\fin
for the off-diagonal case (ii).

For the diagonal model (i) we find $\eta_1 \simeq 10^{-8}$, while in 
the off-diagonal model (ii) this increases enormusly, to  $\eta_1 \simeq 10^{-4}$. This 
suggests that potentially the off-diagonal model is able to provide a 
sufficient amount of asymmetry, looking for appropriate complex models of $M_D$. 

Following this guide-line, we do consider a model of the Dirac 
mass matrix with $i$ factors in several matrix elements. Of course, we first have to check that the
effective neutrino mass matrix is in agreement with the phenomenologically successful neutrino mass matrix (7), and then calculate
the baryon asymmetry. We find that adding $i$ factors in 
positions 2-2, 2-3 and 3-2 this is viable. Therefore, our ansatz for the Dirac neutrino mass matrix is

\ini
M_D \simeq
\left( \begin{array}{ccc}
0 & i \ep^3 & \ep^{4} \\
i \ep^3 & i \ep^2 & i \ep^2 \\
\ep^{4} & i \ep^2 & 1
\end{array} \right)m_t
\fin

We turn now to the schemes (i) and (ii) for the heavy right-handed neutrino masses.
Let us consider first the diagonal model (i).  We find the neutrino mass matrix

\ini
M_{\nu} \simeq
\left( \begin{array}{ccc}
-\ep^6/b+\ep^8 & -\ep^5/b+ i \ep^6 & -\ep^5/b+\ep^4 \\
{*} & -\ep^6/a-\ep^4/b-\ep^4 & i \ep^7/a-\ep^4/b+i \ep^2 \\
{*} & {*} & \ep^8/a-\ep^4/b+1
\end{array} \right) \frac{m_t^2}{m_R}.
\fin

Notice that using the values $b \sim \ep^4$ and $a \sim \ep^5$ proposed in Section IV to get a structure similar to (7), we have lost this structure adopting now the new form for $M_D$ (27), since the (3,3) entry in (28) is small, of the order $\ep^3$. Therefore, maximal mixing can only be achieved in this case with some fine tunning. Moreover, ones finds in this case a too small value of CP violation and baryon asymmetry, very close to (17) and (19).\par

Let us turn now to the more interesting off-diagonal model (ii). We find the neutrino mass matrix

\ini
M_{\nu} \simeq
\left( \begin{array}{ccc}
-\ep^6/c & i \ep^7 -\ep^5/c & \ep^8 -\ep^5/c \\
{*} & -\ep^5-\ep^4/c-\ep^5 & i \ep^6-\ep^4/c+i \ep^3 \\
{*} & {*} & \ep^4-\ep^4/c+\ep^4
\end{array} \right) \frac{m_t^2}{m_R}.
\fin

With $c \sim \ep^2$, as proposed in Section V,  one gets at leading order in $\ep$ exactly the desired form (7) with $m_\nu \sim \ep^2 (m_t^2/m_R)$, that suggests the intermediate scale (22). Moreover, one finds a larger amount of CP violation

\ini
\ep_1 \simeq 4 \cdot 10^{-7}
\fin
In this last case we have 

\ini
\tilde{m}_2=(M_D^{\dg} M_D)_{22}/M_2 \simeq m_{\nu},
\fin
a value that lies in the strong washout regime, so that 

\ini
Y_B \sim 5  \cdot 10^{-12}
\fin
that is somewhat short of the order of magnitude (12).
Remember that in the off-diagonal case the lightest right-handed neutrino
corresponds to the second one in flavor.\par

However, we stress here that for 

\ini
c \sim \ep^{3/2}
\fin 
which is admissible, agreement with leptogenesis is improved, since now 
we have $m_\nu \sim \ep^{5/2}~ (m_t^2/m_R) \sim 0.05$ and we get the scale

\ini
m_R \sim 3 \cdot 10^{11}~GeV,  
\fin 
the CP asymmetry
\ini
\epsilon_1 \simeq 2 \cdot 10^{-6}
\fin
and therefore

\ini
Y_B \sim 1  \cdot 10^{-10}
\fin
that is of the right order of magnitude (12).\par
We have diagonalized the neutrino mass matrix in this latter case (31), and have checked that, to a good approximation, it gives the spectrum and mixing that follow from the simple ansatz (7).

To summarize, in the present context the off-diagonal model at the 
intermediate scale is preferred by leptogenesis with respect to the 
diagonal model at the unification scale.

\section{Conclusion}

The present work relies on the phenomenological approach of textures. It is relevant to recall here that, for example within SO(10), it is possible to obtain any relation between quark and lepton mass matrices by using general Higgs representations, i.e. several {\bf 10}s and {\bf 126}s.
However, we stress that the preservation of the hierarchical structure of mass matrices seems to point towards horizontal symmetries. Then it is also possible to generate relations between mass matrices \cite{dfh}.\par 
In conclusion, we propose to relax the simple 
quark-lepton symmetry for the Dirac neutrino mass matrix $M_D = M_u$ and consider 
a kind of generalized quark-lepton symmetry that keeps the hierarchy of the matrix elements
in terms of powers of $\ep$. We could call this phenomenological approach $hierarchical$ quark-lepton symmetry. In particular, we have considered in detail a case in which the moduli of the matrix elements of $M_D$ are equal to those of $M_u$.\par 
Out of the several models, we have found one scheme 
consistent with the neutrino mass spectrum, lepton mixing and leptogenesis:
the off-diagonal matrix (6) with one lightest
right-handed neutrino corresponding to the second flavor, together with a 
Dirac mass matrix (29) with imaginary second row and column. The diagonal form 
is known to be viable only for degenerate lightest masses of the Majorana heavy
neutrinos.

\begin{appendix}

\begin{center} 
\vskip 0.5 truecm  {\bf APPENDIX}\par \vskip 3 truemm
\end{center}

We consider here two other interesting possibilities of rank-3 matrices for the heavy Majorana neutrinos :

$$M_R =
\left( \begin{array}{ccc}
d & 0 & 0 \\
0 & 0 & 1 \\
0 & 1 & 0
\end{array} \right)m_R,\eqno({\rm A.1})$$

$$M_R =
\left( \begin{array}{ccc}
0 & e & 0 \\
e & 0 & 0 \\
0 & 0 & 1
\end{array} \right)m_R.\eqno({\rm A.2})$$

For these two models, the effective neutrino mass matrix is not 
in agreement with the structure (7). However, considering the real case, 
since the structure in sector 2-3 of the effective matrix is achieved and 
the smallness of the first row and column could be due to running effects 
\cite{smir}, we perform again the calculation.
In model (A.1) we get $d \sim \ep^4$ and
$m_R \sim 10^{13}$ GeV and, moreover, $Y_L \sim 10^{-13}$. In model (A.2) we 
obtain $e \sim \ep^5$ and $m_R \sim 10^{16}$ GeV and the lepton asymmetry is 
enhanced due to degeneracy in the two lightest right-handed neutrinos 
\cite{ab}.

\end{appendix}

\newpage

\begin{center} 
\vskip 0.5 truecm  {\bf ACKNOWLEDGEMENTS}\par \vskip 3 truemm
\end{center}

One of us (L.O.) acknowledges partial support by the EU Contract No. MRTN-CT-2006-035482 (FLAVIANET) and an advise by Jean-Claude Raynal.

\end{document}